\documentclass[twocolumn,showpacs,superscriptaddress,preprintnumbers,amsmath,amssymb,pra]{revtex4}
\usepackage{}

\usepackage{mathrsfs}
\usepackage{amsmath}
\usepackage{amssymb}
\usepackage{graphicx}
\usepackage{dcolumn}
\usepackage{bm}
\usepackage{subfigure}
\usepackage{color}
\usepackage{ifpdf}
\usepackage{epstopdf}
\usepackage{CJK}
\usepackage{float}

\begin{document}

\title{Long-time nonlinear dynamical evolution for P-band ultracold atoms in an optical lattice}

\author{Dong Hu}
\affiliation{School of Electronics Engineering and Computer Science, Peking University, Beijing 100871, China}
\author{Linxiao Niu}
\affiliation{School of Electronics Engineering and Computer Science, Peking University, Beijing 100871, China}
\author{Baoguo Yang}
\affiliation{School of Electronics Engineering and Computer Science, Peking University, Beijing 100871, China}
\author{Xuzong Chen}
\affiliation{School of Electronics Engineering and Computer Science, Peking University, Beijing 100871, China}
\author{Biao Wu}
\affiliation{International Center for Quantum Materials, School of Physics, Peking University, Beijing 100871, China}
\affiliation{Collaborative Innovation Center of Quantum Matter, Beijing 100871, China}
\affiliation{Wilczek Quantum Center, College of Science, Zhejiang University of Technology,  Hangzhou 310014, China}
\author{Hongwei Xiong}
\affiliation{Wilczek Quantum Center, College of Science, Zhejiang University of Technology,  Hangzhou 310014, China}
\author{Xiaoji Zhou}\email{xjzhou@pku.edu.cn}
\affiliation{School of Electronics Engineering and Computer Science, Peking University, Beijing 100871, China}
\date{\today}

\begin{abstract}
We report the long-time nonlinear dynamical evolution of ultracold
atomic gases in the P-band of an optical lattice. A Bose-Einstein
condensate (BEC) is fast and efficiently loaded into the P-band at
zero quasi-momentum with a non-adiabatic shortcut method. For the
first one and half milliseconds, these momentum states undergo
oscillations due to coherent superposition of different bands, which
are followed by oscillations up to 60ms of a much longer period. Our
analysis shows the dephasing from the nonlinear interaction is very
conducive to the long-period oscillations induced by the variable
force due to the harmonic confinement.
\end{abstract}

\pacs{67.85.-d;03.75.Lm;03.75.Hh; 37.10.Jk}

\maketitle

\section{introduction}
The occupation of high orbital quantum state in solid-state system
is known to play a crucial role to the understanding of a lot of
strongly correlated systems, such as high-temperature
superconductivity and semiconductor~\cite{superc}. Ultracold atoms
in an optical lattice have been used to simulate quantum many-body
systems in the ground state~\cite{Bloch}. Recently, there has been
increasing efforts to study ultracold atomic gases in optical
lattices at high orbitals~\cite{Liu1}. This kind of study can not
only help to understand high orbital quantum state in solid-state
system but also provide a chance to study new physics which is
completely beyond that of the solid-state system, such as the
emergence of the quantum coherence in the first excited Bloch band
of an optical lattice~\cite{Bloch2}. High orbital physics in cold
atomic gas has attracted a lot of attention; there are many
theoretical predictions, such as supersolids~\cite{Ssolid} and other
novel phases~\cite{Np1,Np2,Larson}.

The long time dynamics of a Bose-Einstein condensate (BEC) in the quantum state of the S-band has been intensively
studied and many interesting phenomena, such as coherent oscillations \cite{ScienceOs,BO}
and nonlinear self-trapping \cite{selfT1,selfT2} were observed.
However, it is much more challenging to observe the long-time dynamics of a BEC
in high orbitals because of the difficulty to prepare the cold atoms in such states and maintain their
quantum coherence. For the condensate with a wide quasi-momentum distribution in P-band~\cite{Bloch2},
the quantum dynamics for the holding time being shorter than 1.3 ms were studied.

Here we experimentally prepare a BEC in the P-band of an optical
lattice using a non-adiabatic shortcut method and study its long time quantum dynamics.
The initial BEC in the P-band with finite size at quasi-momentum $q=0$ has a local maximum in energy so that
this state is unstable when the harmonic confinement and interatomic interaction are considered.
 For the first one and a half milliseconds, the momentum states undergo fast oscillations.
After a brief intermediate interval, the BEC starts a different type of oscillations which
have a much longer period ($\sim$ 14.9ms) and last up to 60ms. We find that these long-period oscillations are due to
the presence of an additional harmonic trap, and the dephasing induced by
the nonlinear interaction is greatly helpful for this oscillation behavior.
This counterintuitive features and long time quantum coherence are possible to study the quantum thermalization.

\section{ Preparation of P-band quantum state with a shortcut method}
In our experiment,
we first prepare a pure BEC of about $2\times10^5$ $^{87}$Rb
atoms in a hybrid trap which is formed by overlapping a single-beam
optical dipole trap with wave length 1064nm and a quadrapole
magnetic trap. The resulting potential has harmonic trapping
frequencies $(\omega_x,\omega_y,\omega_z)=2\pi\times(28,55,65)$Hz,
respectively. Our lattice is produced by a standing wave created by
two counter-propagating laser beams with lattice constant
$a=\lambda/2=426$ nm along the $x$-direction.

\begin{figure}
\begin{center}
 \includegraphics[width=0.45\textwidth]{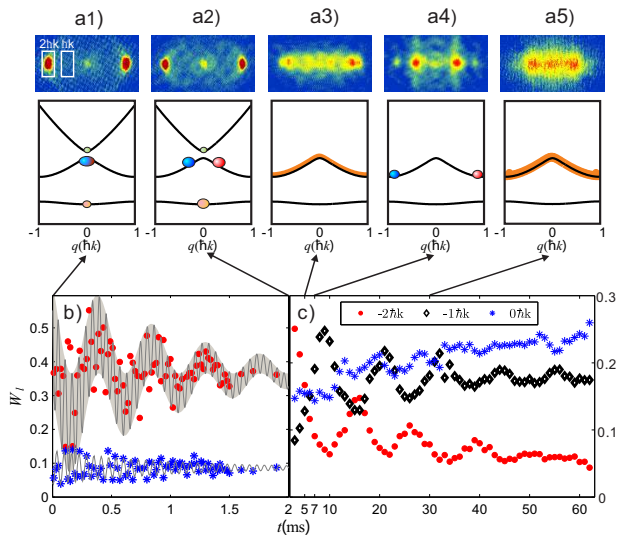}
\end{center}
\caption{(a1-a5) The first row: the measured momentum distributions
of the BEC in the P-band at holding different times (0ms, 2ms, 5ms, 7ms, 30ms), and the white rectangles are region for us to calculate proportion of different momentum states; the second row: schematic illustration of the corresponding population distributions in the Bloch band. (b) Population oscillations around momenta $0\hbar k$ (blue stars) and $-2\hbar k$ (red dots) with $t< 1.5$ms. Solid lines are from numerical simulations. The shade is to highlight the slow oscillating envelop of the rapid oscillations. (c) Population oscillations around momenta $0\hbar k$ (blue stars), $-\hbar k$ (black diamonds), and $-2\hbar k$ (red dots) with $t>2$ms.
The oscillation period is about 14.9ms with $V_{0}=5 E_{r}$.}
\label{figure2}
\end{figure}

We then load the BEC into the P-band by applying a series of pulsed optical lattices within tens of microseconds~\cite{Zhai,Liu,Chen,Campo}. There are many other methods preparing a BEC in the P-band quantum state~\cite{Bloch2,NP,speed,shaken,GW}, but normally the loading time is tens of milliseconds.
Our method has the merit of rapid generation of
the quantum state in the P-band. More importantly, this method allows us
to generate the desired P-band with very narrow quasi-momentum width around $q=0$, which is crucial to observe the long-time
quantum dynamics. Most recently, this method is
proposed to study anisotropic two-dimensional Bose
gas~\cite{Papoular}.

We use two acousto-optic modulators to form our
designed pulse sequences with the frequency difference
$\delta\omega=182.5$MHz which corresponds to a phase shift between
two pulses series by $3\pi/4$. This phase shift is crucial to change the parity of the ground state and thus prepare the state in the P-band. Four special pulses with a chosen depth $V_0$ are used to prepare the state in the P-band around $q=0$. The details of the pulse series are shown in the Supplementary Material.

For the quantum state in the S-band with $q=0$, the momentum
distribution has peaks around $0\hbar k$ and $\pm 2\hbar k$ with $k=2\pi/\lambda$
\cite{lattice,Liu}. For the quantum sate in the P-band with $q=0$,
because its wave function has the odd parity
$\psi(-x)=-\psi(x)$, the momentum distribution equals
to zero at $0\hbar k$ and has significant peaks at $\pm 2\hbar k$, as shown in
Fig.~1(a1), where two dominating peaks at $\pm 2\hbar k$
are clearly seen.

\section{ Dynamical evolution of the quantum state in P-band}
To study the dynamical evolution after a BEC is prepared in the quantum state
of the P-band, we hold the condensate for time $t$ before measuring the momentum distribution of the ultracold atoms after a time of flight (TOF) of 28ms. The momentum distributions of the BEC at $t=$0ms, 2ms, 5ms, 7ms, 30ms are shown in the first row of
Fig.~1(a1, a2, a3, a4, a5). An evolution is clearly observed. The corresponding
evolution in the Bloch bands is schematically illustrated in the second row.
The depth of the optical lattice is $V_0=5E_r$ with $E_r$ being the atomic recoil energy.

The momentum distributions are analyzed by computing the normalized atom populations
around momentum states $|l\hbar k\rangle$ ($l=0,-1,-2$) as $W_{l}(t)=|\langle l\hbar k|\psi(t)\rangle|^2=N_{l}(t)/{N}$. Here $N$ is the total atom number and $N_{l}(t)$ is the atomic
number around $|l\hbar k\rangle$. In the TOF images we choose $95.2\mu$m as
the width to determine the atom number $N_l$. This is the width of a condensate
without any quantum manipulation and after the free expansion of the same TOF.
We find $W_{l}$'s oscillation with time as shown in Fig.~1(b, c).
In Fig.~1(b), there are rapid oscillations in $W_{-2}(t)$ and $W_0(t)$.
These rapid oscillations disappear around $t=1.5$ms. After a short transition time,
a different type of oscillations begin to emerge around $t=$2ms. These oscillations
can be observed up to 60ms and have a much longer period of 14.9ms.
We will show how the interaction would affect this long-period oscillation, which was not observed in the previous work~\cite{Bloch2}.

\section{ Short-period oscillations}
The short-period oscillations shown in Fig.~1(b) are beating signal due to coherent superposition of
different bands. In real experiment, it is impossible to prepare a
quantum state completely in the P-band. What we prepared is in fact
a superposition of s, p and d bands with negligible higher
band. Due to difference in timescale of two oscillations and
uncertainty in experiment, the experimental data is shown in the
figure comparing with the peripheral contour of the beating signal
from theoretical simulation with the choice of the initial quantum
state $|\psi(t=0)\rangle=\sqrt{0.9}|\mathrm{p},q=0\rangle+
\sqrt{0.05}|\mathrm{d},q=0\rangle+\sqrt{0.05}|\mathrm{s},q=0\rangle$.

With this understanding, the periods of the oscillations in
Fig.~1(b) are determined by the band gaps. As there are three band
gaps between the s, p, and d bands, there should be three
oscillating periods, $T_{\mathrm{sp}}$, $T_{\mathrm{sd}}$, and $T_{\mathrm{pd}}$. Around
$0\hbar k$, the atoms are mostly from the s and d bands due to
the odd parity of the P-band. As a result, we expect only one
oscillation period $T_{\mathrm{sd}}=60.0\mu$s. This is indeed what we see for $W_0$ in
Fig.~1(b). Around $-2\hbar k$, all three bands have significant
contributions. Because the populations of the s and d bands are
both small, the contribution to $W_{-2}$ related to period $T_{\mathrm{sd}}$
is of second-order and negligible. As a result, only two periods are
expected for $W_{-2}$, $T_{\mathrm{sp}}$ and $T_{\mathrm{pd}}$. The oscillations of
$W_{-2}$ in Fig.~1(b) clearly have two different periods, rapid
oscillations enveloped by slow oscillations. This is consistent with
the fact that $T_{\mathrm{pd}}$ is much larger than $T_{\mathrm{sp}}$. For $V_0=5 E_r$
we have $T_{\mathrm{sp}}=69.5\mu$s and $T_{\mathrm{pd}}=465.5\mu$s.

The short-period oscillations in Fig.~1(b) have a slow decay. This
decay is mainly due to the decay of population on P-band induced by collision between atoms~\cite{Zhai},
which is already considered in theoretical results in Fig.~1(b)
(solid lines). For our system with atomic number $2\times10^5$ and
density of $6.4\times10^{13}$cm$^{-3}$ for $V_0=5 E_r$, the
calculated decay time is $1.3$ms. It is coincident to the fitting
result ($1$ms) based on the experimental data, and the decay rate in experiment is larger which can be attributed to other mechanism including dephasing of the atom cloud. Considering its finite
size, the condensate has a narrow quasi-momentum distribution around
$q=0$ with a width of $0.06\hbar k$. Our numerical simulation finds
that the calculated decay time would increase by about $1.9\%$, and
numerical simulation shows the influence of this distribution to
oscillation frequency is also negligible.

\section{Long-period oscillations in the P-band}The rapid
oscillations disappear at about $1.5$ms. After a featureless short
dynamical evolution, a different type of oscillations begin to
emerge at around 2ms. As shown in Fig.~1(c), these oscillations have
a much longer period of about $15$ms and their oscillating amplitude
subsides gradually and becomes zero around $60$ms after five cycles.
This is different to the oscillation period of the harmonic trap
$35$ms. If we consider this evolution based on the semi-classical
model with an effective potential, where the atoms exerted a
variable force due to the harmonic trap and quasi-momentum dependent
effective mass, as shown in the Supplementary Material, we see that
the period is identical to the experimental result. Different to the
Bloch oscillation for atoms on S-band~\cite{BO,Gustavsson,Kling},
however, our analysis shows that these long-period oscillations for
atoms on P-band are greatly affected by the nonlinear dephasing, and
this semi-classical model can not fully coincident to the
experimental observation.

\begin{figure}
\begin{center}
 \includegraphics[width=0.45\textwidth]{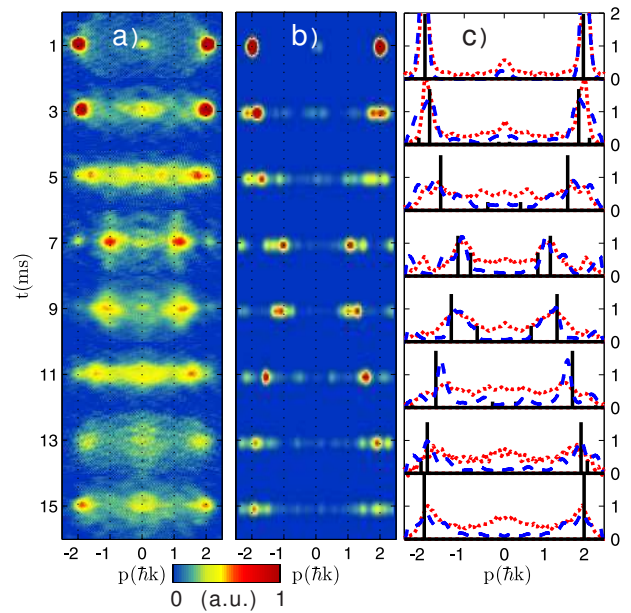}
\end{center}
\caption{The two-dimensional momentum distributions of the first
cycle are shown in (a) and (b) for experiment and
theoretical simulation, respectively. After an integration along the
vertical direction, (c) gives the one-dimensional distribution for experiment (red dotted line) and numerical
simulation (blue dashed line). The black line is obtained
from the semiclassical model without considering interatomic
interaction.}.
 \label{figure4}
\end{figure}

We now look at the experiments in detail. Shown in Fig.~\ref{figure4}(a)
are the momentum distributions measured experimentally for holding
times up to 15ms, which is roughly the first cycle of
the oscillations.  At $t=0 $ms, the momentum mainly distributes
around $\pm 2\hbar k$ with $q=0$, which is the signature
of the initial state in the P-band. At $t=3 $ms, the
absolute value of the peak quasi-momentum becomes smaller
than $2\hbar k$. This is due to that the kinetic energy is transferred
into the potential energy by the harmonic trap.
At $t=5$ms, the atoms spread to the full regime
between $-2\hbar k$ and $+ 2\hbar k$. At $t=7$ms,
the atoms concentrate around $\pm \hbar k$, an indication of the quantum
state has moved to the Brillouin zone edge of the P-band (Fig.~1(a4)).
Around $15$ms, most of the atoms have moved back to positions around $\pm 2\hbar k$.
Overall, an oscillating pattern is clearly observed in these momentum distributions.

The oscillation pattern can be well captured by our simulation with
the Gross-Pitaevskii equation (GPE) as seen in Fig.~\ref{figure4}(b), where the simulation is to begin with the ground state
in harmonic trap, and carry out with the same experimental process. At the center of the momentum distribution, the experiment shows significant population, which is due to the presence of incoherent atoms.
To show it more clearly, we have integrated out the vertical direction and compared the experimental and numerical results in Fig.~\ref{figure4}(c). We see that the peak positions
are seen clearly matching very well. The experimental results (red
dotted line) agree with the theoretical simulation (blue
dashed line), where the black line is based on the semi-classical model.

\section{ Effects of the random phase between neighboring lattice sites on the long-period oscillation}
\begin{figure}
\begin{center}
 \includegraphics[width=0.45\textwidth]{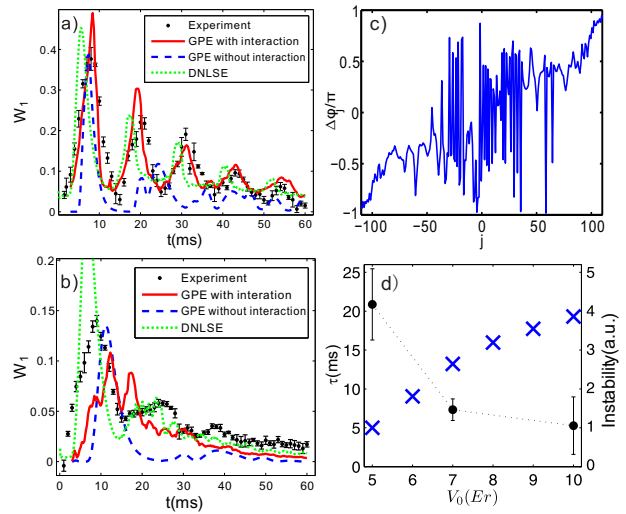}
\end{center}
\caption{The population with $\hbar k$ versus time $t$ for $5E_{r}$
(a) and $10E_{r}$ (b). The experimental results, theoretical
simulations with and without the interaction correspond to the black
dots with error bar, red solid curves and blue dashed curves,
respectively, and the green dotted curves are the result of DNLSE.
(c) The phase differences between neighboring lattice site $j$ for
$V_{0}=5E_r$ and $t=5$ms. (d) The dotted points are the measured
decay time of the oscillation amplitude, and the cross points are
for the dynamical instability ratio.}
 \label{figure5}
\end{figure}

The observed time evolution (black dots with error bar) of the proportion with momentum $\hbar k$ is given in
Fig.~\ref{figure5}(a) for $V_{0}=5E_{r}$, where
we can distinguish five cycles of oscillation. In the figure we have extracted incoherent atoms at the center, and correspondingly the numerical simulation is multiplied by a decay factor. For $5E_{r}$, the theoretical simulation with interatomic
interaction (red solid curve) agrees perfectly with the experimental result.
It is quite surprising that the
simulation without interatomic interaction (blue dashed curve) doesn't show clear periodic behavior, which differs significantly with the experiment.
Both in
experiment and theoretical simulation, the interaction has an effect
of stabilizing the oscillation. This is counterintuitive by noticing
that (i) the repulsive interaction has the effect of widening the wave
packet and thus it seems that it should destroy the oscillation; (ii) the interaction usually plays a role of destroying the quantum coherence because of the nonlinear interaction.

To solve this puzzle, we consider the following evolution of the
order parameter of the condensate,

\begin{eqnarray}\label{eqn:disGPE}
\Psi (x,t) = \sum_j \sqrt{N_j} \psi_j(x,t).
\end{eqnarray}
Here $N_j$ is the number of atoms initially in the $j$th lattice site, and $\psi_j(x,t)=\sum_n \alpha_{jn}(t) W_n(x)$ with $W_n(x)$ the Wannier wave function located in the
$n$th lattice site and $\alpha_{jn}(t)$ the corresponding coefficient. For atoms in $j$th site, initial situation is
$\alpha_{jn}(0)=\delta_{jn}$. During evolution of the single site wave
function, population would transfer from the $j$th site to different
site $n$, and $\alpha_{jn}(t)$ can be rewritten as
$|\alpha_{jn}(t)| e^{\phi_{jn}(t)}$. Despite a smoothly
varying of the external harmonic potential, the contribution of both
the kinetic energy and potential energy to the phase $\phi_{jn}(t)$
is the same for every lattice site. However, the phase due to the
interatomic interaction is proportional to each site's atom number
which could be quite different for neighboring lattice sites. As
shown in Fig.~\ref{figure5}(c), the phase difference between
neighboring lattice sites appears completely random because of
interatomic interaction at $t=5$ms. In a sense, the relative phase
between neighboring sites is pseudorandom because it appears random
only after the relative phase is confined between $-\pi$ and $\pi$.

In this case, when the density distribution or momentum distribution
is considered, the interference term between different
$\alpha_{jn}(t)$ could be omitted, i.e.
\begin{eqnarray}\label{eqn:disGPE}
\nonumber |\Psi(x,t)|^2&\approx&\sum_j|\psi_j(x,t)|^2\\
&=&\sum_j N_j \left| \sum_n \alpha_{jn}(t) W_n(x)\right|^2.
\end{eqnarray}
With the above approximate expression, we solve every $\psi_j(x,t)$
by the discrete nonlinear Schr\H{o}dinger equation (DNLSE)
\cite{ScienceOs}, and then get the momentum distribution, as shown
with the green dotted lines in Figs.~\ref{figure5}(a) and
\ref{figure5}(b). In this situation, our numerical calculations show
periodic behavior with each peak shift ahead a little comparing with
experiment, however, the period of the time agrees very well with
the experimental result. In the Supplementary Material, we give the
evolution of the $|\psi_j(x,t)|^2$ which shows that there is long
time period behavior in $|\psi_j(x,t)|^2$. For the condensate in the
P-band, because in every $\psi_j(x,t)$, it has the peaks around
$-2\hbar k$ and $2\hbar k$, the wave packet in every lattice site
will split into two parts, propagate in different directions and
oscillate due to the presence of the harmonic trap. It is natural
that the incoherent superposition in Eq.(\ref{eqn:disGPE}) will also
have the long time periodic oscillation. For $V_0=5E_r$ the GPE
agrees better with the experiment than the method of DNLSE and
Eq.(\ref{eqn:disGPE}), while here the DNLSE simulation shows the
physical picture more clearly.

To verify further the role of random phase in the appearance of the long-time period behavior, we add this random phase artificially for noninteracting case and solve numerically the Schr\H{o}dinger equation. We find that there is perfect periodic behavior in $W_1(t)$, although the period is a little different from the interaction case.

In Fig.~\ref{figure5}(b), we give the experimental result for
$V_0=10E_{r}$. We see that the amplitude of the oscillation for
$10E_{r}$ is much smaller than $5E_{r}$. For $10E_{r}$, there is a
much stronger damping which is due to the highly suppressed
tunneling between neighboring sites. In addition, there is
no clear periodic behavior in this case. The simulation
with GPE doesn't agree well with the experimental result.
This may due to the fact that at $10E_{r}$, the particle number is
highly squeezed so that there is a breakdown of the mean-field
theory, which induces the disappearance of the
interference between $\psi_j(x,t)$. It seems that the theoretical simulation with GPE for $5E_r$
agrees better with the experiment than the method with DNLSE.

\section{ Discussion and Summary}
In the long time evolution, the
oscillation amplitude and damping are greatly affected by the
dynamical instability due to the interaction between atoms. We
calculated the dynamical instability~\cite{insta,insta2} of P-band
at lattice depth varying form 5 to 10$E_r$. In Fig.~\ref{figure5}(d)
the cross points show the calculated dynamical instability with the
ratio of different lattice depth and 5$E_r$. The solid circles give
the lifetime of oscillation by fitting the experimental data with
exponential decay. As expected, the increasing of $V_{0}$ will lead
to larger dynamical instability ratio, and lead to the suppression
of the oscillation.


In summary, we demonstrate an effective way of loading atoms
directly into P-band using two group of lattice pulses with
different phases to break the parity of the Hamiltonian. Our
preparation of the P-band with $q=0$ is different from
Ref.~\cite{Bloch2} where the condensate has a distribution in the
regime between $q=-\hbar k$ and $\hbar k$. In our work, the
evolution of the condensate in the P-band is studied experimentally.
The short-period oscillation due to coherent superposition of
different bands and long-period oscillation reflecting the random
relative phase between neighboring lattice sites are observed
simultaneously. The present experiment paves the way to study the
long time dynamical evolution of the high orbital physics for other
novel quantum state such as fermonic superfluid, molecular
condensate and condensate with special configuration of optical
lattice.

\section*{ACKNOWLEDGEMENT}

This work is partially supported by the state Key
Development Program for Basic Research of China No.2011CB921501, NSFC (Grants No.61475007, No.11334001 and No.91336103),
RFDP (Grants No.20120001110091).

\appendix{}
\section{Loading method}

The Hamiltonian of a single-atom in a one-dimensional optical
lattice can be expressed as:
\begin{eqnarray}\label{eqn:H_single}
H_{x}={p_{x}^{2}}/{2m}+V_{0}\cos ^{2}\left(kx\right),
\end{eqnarray}
where $V_{0}$ is the depth of the optical lattice, which is in unit
of $E_{r}$, $E_{r}={(\hbar k)^2}/{2m}$ is the recoil energy of the
atom, $p_{x}$ is the momentum of the atom along $x$-direction, $m$
is the atom mass, $k=\pi/a$ is the wave vector and $a$ is the
lattice constant. The eigenstate of the Hamiltonian
Eq.~(\ref{eqn:H_single}) is the Bloch state $|n,q\rangle$.

\begin{figure}
\begin{center}
\includegraphics[width=0.45\textwidth]{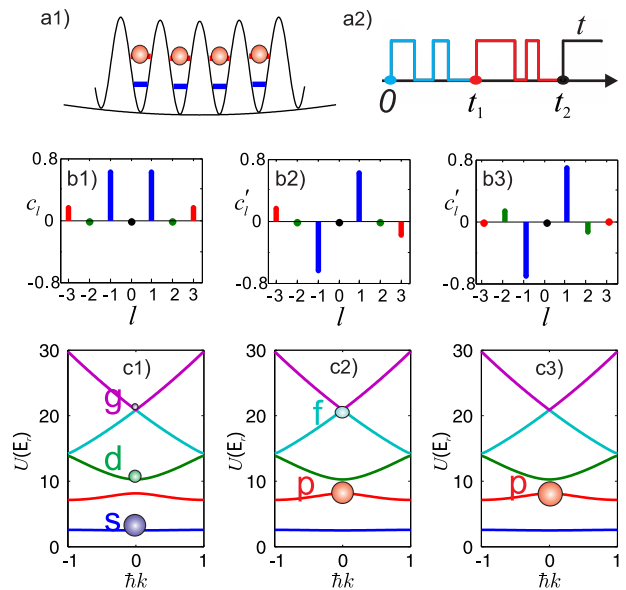}
\end{center}
\caption{The shortcut loading process of the atoms into P-band in 1d
optical lattice. ($a1$) The schematic diagram of atoms in P-band.
($a2$) Loading time sequences with a different phase between the
first series of pulses from $0$ to $t_{1}$ and the second from
$t_{1}$ to $t_{2}$. $t$ is the holding time in the optical lattice.
($b1$--$b3$) illustrate the superposition coefficient $c_{l}$ right
before $t_1$ and $c^\prime_{l}$ at the moment right after $t_1$ and
at $t_2$, respectively. (c1-c3) illustrate the corresponding
population distribution in the Bloch band.} \label{fig4}
\end{figure}

Since the system is periodic, and our loading method conserves the
quasi-momentum, we can choose the plane waves $|2l\hbar k+q\rangle$
with quasi-momentum $q$ as a set of complete bases, i.e.,
\begin{eqnarray}\label{eqn:psi_plane}
|\psi\rangle=\sum_{l} c_{l} |2l\hbar k+q\rangle,
\end{eqnarray}
where $c_{l}$ is the superposition coefficient, $l=0,\pm 1,\pm 2...$
is an index number, $q\in \lbrack -k,k)$ is the quasi-momentum, and
$|2l\hbar k+q\rangle\equiv e^{i(2lk+q/\hbar)x}/\sqrt{2\pi}$ is the
eigenstate of $p_x$. On the other hand, the state $|\psi\rangle$ can
also be expressed in the Bloch state $|n,q\rangle$ as
$|\psi\rangle=\sum_{n} \langle n,q| \psi\rangle |n,q\rangle$, where
$n=s,p,d...$ is the band index. The aim state $|\psi_{a}\rangle$ is
the first excited state at $q=0$ , as shown in
Fig.~\ref{fig4}($a1$).

Our shortcut method includes a series of designed standing wave
pulses. The atom state $|\psi\rangle = \prod\limits_{j = 1}^4
{U_{f_{j}}(t_{f_{j}})\cdot
U_{p_{j}}(t_{p_{j}})\cdot|\psi_{0}\rangle}$, where
$U_{p_{j}}(t_{p_{j} })$ and $U_{f_{j}}(t_{f_{j}})$ are the time
evolution operator and free evolution operator for the $j$th pulse
with duration $t_{p_{j}}$ and $t_{f_{j}}$, respectively. The atom
experiences spatial period potential in the pulse duration, and the
different momentum states can gain different phase factors between
the interval.

Our preparation process consists of two series of pulses as shown in
Fig.~\ref{fig4}($a2$), where we need optimize the fidelity
$|\langle\psi_a|\psi\rangle|^2$ between the aim state
$|\psi_a\rangle$ and $|\psi\rangle$ to its maximum by changing the
parameters of pulse sequence. In the first series of pulses from $0$
to $t_{1}$, the atom experiences spatial potential $V_0\cos
^{2}\left( x\pi /a\right)$. For the second series pulses from
$t_{1}$ to $t_{2}$ the atom experiences potential $V_0\cos^{2}\left(
x\pi/a+3\pi/4\right)$. The coefficients $c_{l}$ ($c^\prime_{l}$) and
the distribution in the Bloch band are shown in Fig.~\ref{fig4}(b)
and (c), respectively.

For a pure Bloch state or a superposition state with different Bloch
state at $q=0$, the parity can be given as $P=\sum\limits_{l}
|c_l-c_{-l}|^2/4$, where $P=1$ stands for a state with odd parity
and $P=0$ even.

At time $t_{1}$, all the components in the parity $P$ would satisfy
$c_l-c_{-l}=0$, as shown in Fig.~\ref{fig4}($b1$), and the atom is
distributed in the s,d,g... bands, as in Fig.~\ref{fig4}($c1$).
However, from the view of the second series of pulses, by the
lattice shift, the coefficient $c_l$ would be appended with a phase
according to $l$, as
$c^\prime_l|l\rangle=c_le^{i2l(3\pi/4)}|l\rangle$, and the relation
between the coefficients become
$c^\prime_l-(-1)^{l}c^\prime_{-l}=0$. In our loading process, the
first series of pulses ensure that coefficient $c^\prime_l$ with
even $l$ is zero, and thus the parity of state can be completely
changed as shown in Fig.~\ref{fig4}($b2$). Its energy band
distribution is in the p,f,.., as drawn in Fig.~\ref{fig4}($c2$).
At time $t_{2}$, as shown in Figs.~\ref{fig4}($b3$) and ($c3$), with
another two pulses conserving the parity, only P-band state is
populated.

\section{Effective potential model for p band dynamics}
\begin{figure}
\begin{center}
\includegraphics[width=0.45\textwidth]{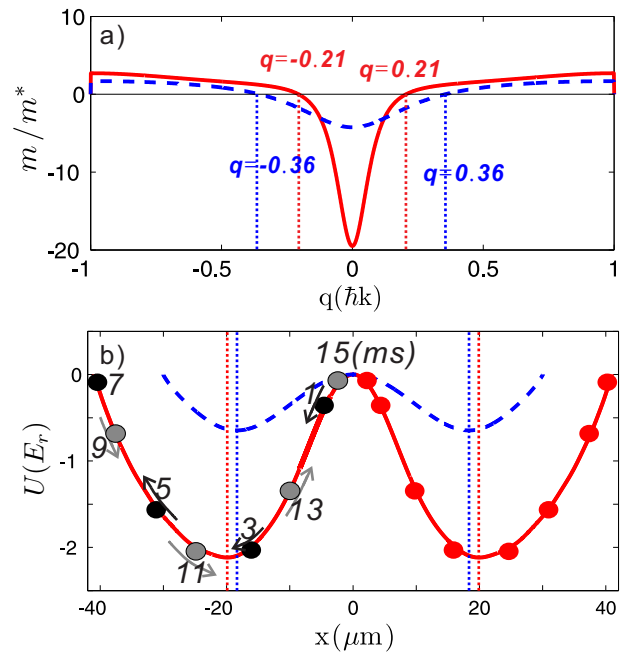}
\end{center}
\caption{The effective mass for atoms in P-band (a) and its
effective potential in the presence of a harmonic trap (b) for
$5E_{r}$(red solid line) and $10E_{r}$(blue dashed line). The dotted
lines correspond to the points where the sign of the effective mass
changes.} \label{fig5}
\end{figure}

Here we consider a semi-classical model to understand the periodic
dynamics in the P-band. In the optical lattice, the effective mass
is $m^{*}(q)=\hbar (\partial v/\partial
q)^{-1}=\hbar^{2}(\partial^2E_{p}(q)/\partial q^2)^{-1}$, where $v$
is the group velocity and $E_{p}(q)$ is the energy of the P-state
for different $q$. In Fig.~\ref{fig5}(a), the ratio between $m$ and
$m^*$ is shown for $V_{0}=5E_{r}$ and $10E_r$, respectively. We see
that the effective mass changes sign at the $q=\pm 0.21\hbar k$ and
$\pm 0.36 \hbar k$ for $V_{0}=5E_{r}$ and $10E_r$, respectively.

With the semi-classical model, we simplify the problem by
considering a quasi-particle with $q=0$  and study the dynamical
process in the harmonic trap. When the effective mass is considered,
the acceleration is given as $a(x)=\frac{F(x)}{m^*}$~\cite{mass},
where $F(x)=-\partial(\frac{1}{2} m \omega^2 x^2)/\partial
x=-m\omega^2 x$ is the force driven by the harmonic trap without the
optical lattice. Using $\frac{dv}{dt}=a$, $\frac{dx}{dt}=v$,
$\frac{dq}{dt}=F$ and the initial condition, we get $q$, $m^*$, $x$
and $F$ at time $t$. Then by expressing all the parameters as
function of $x$, the dynamical evolution can be described by an
effective potential:
\begin{eqnarray}\label{eqn:mass}
U(x)=-\int_0^{x}\frac{m}{m^*(x^\prime)}F(x^\prime)dx^\prime.
\end{eqnarray}
In Fig.~\ref{fig5}(b), we give the numerical results of the
effective potential. This means that the external confinement acts
as a repulsive potential~\cite{mass,massPu} and the atoms at trap
center would be divided into two symmetrical parts for the different
momentum states. The minimums of the potential correspond to the
points where the sign of the effective mass changes from the
negative to positive. The initial atoms in the P-band at $q=0$ and
$x=\pm 5\mu$m near the center of the effective potential, would move
from the center to the edge (as points denoted by time $1,3,5,7$ms)
and then come back (as points by time $9,11,13,15 $ms) and form an
oscillation.

\section{Simulation with single site DNLSE} 
\begin{figure}
\begin{center}
\includegraphics[width=0.45\textwidth]{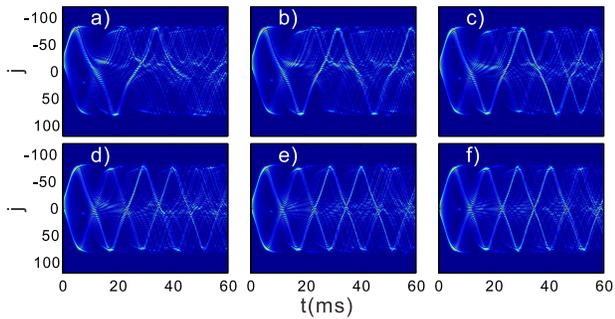}
\end{center}
\caption{Evolution of different site's wave function for $60ms$ with
$5E_r$ lattice depth. The simulation includes totally $301$ lattice
sites. The wave function (a-f) is initially distributed in the
$-25th$, $-20th$, $-15th$, $-10th$, $-5th$ or $0th$ site
respectively.} \label{position5Er}
\end{figure}

\begin{figure}
\begin{center}
\includegraphics[width=0.45\textwidth]{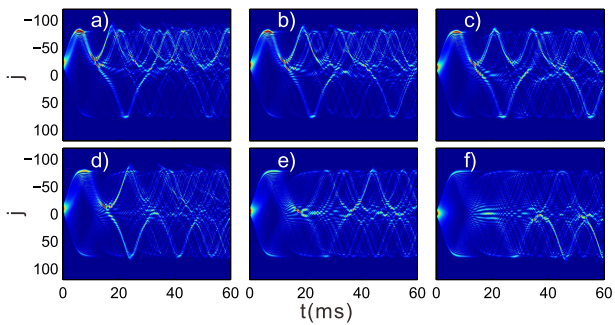}
\end{center}
\caption{Evolution of different site's wave function for $60ms$ with
$10E_r$ lattice depth. Other conditions are the same as Fig.~3.}
\label{position10Er}
\end{figure}

Here we demonstrate the
long-period oscillation from the random phase between neighboring
lattice sites in detail. As shown before, the simulation without
interatomic interaction doesn't have a perfect oscillation. However,
when the interference terms between different sites are omitted, we
can simulate the evolution of each single site based on the
DNLSE~\cite{ScienceOs}, and the population $n_j$ would evolute with
time as shown in Figs.~\ref{position5Er} and \ref{position10Er} for
$5E_r$ and $10E_r$ lattice depth, respectively. For $5E_r$, atoms in
different lattice sites would be distributed more concentrated
during the oscillation, while for $10E_r$ the atoms would disperse
more rapidly into different lattice sites.

According to the evolution of the wave function in the $j$th lattice
site $\psi_j(x,t)$ of the optical lattice and the harmonic trap, for
each site's evolution, there is a perfect oscillation for the ratio
of atoms with momentum $\hbar k$ versus the holding time in the
potential. The initial state $\Psi(x,t=0)$ is the superposition of
Wannier functions in different lattice sites. Considering the fact
that the randomization between neighboring lattice sites arises from
the atomic interaction, the total momentum distribution is given as
the weighted incoherent superposition of every site's wavefunction's
evolution. Therefore, the whole atom cloud could also exhibit a good
oscillation behavior with time. However, if the coherence term
exists, the result would be different, and the collective
oscillation is destroyed as shown by the simulation in Fig.~3 of
main body.

\emph{}

\end{document}